\documentclass[fleqn,usenatbib]{mnras}
\usepackage{amsmath,amstext}
\usepackage{mathptmx}
\usepackage{txfonts}
\DeclareSymbolFont{CMletters}{OML}{cmm}{m}{it}
\DeclareMathSymbol{v}{\mathalpha}{CMletters}{`v}

\usepackage[T1]{fontenc}
\usepackage{ae,aecompl}
\usepackage{graphicx}    

\usepackage[utf8x]{inputenc}
\usepackage[encapsulated]{CJK}  

\newcommand{\cntext}[1]{\begin{CJK}{UTF8}{bkai}#1\ignorespacesafterend\end{CJK}} 

\title[A circumplanetary disc with a dead zone]{The evolution of a circumplanetary disc with a dead zone}
 
\author[Chen et al.]{
Cheng Chen \thanks{E-mail: chenc21@unlv.nevada.edu}, Chao-Chin Yang (\cntext{楊朝欽}), 
Rebecca G. Martin and Zhaohuan Zhu  
\\ Department of Physics and Astronomy,  University of Nevada, Las Vegas, 4505 South Maryland Parkway, Las Vegas, NV 89154, USA 
}
 
\date{Accepted XXX. Received YYY; in original form ZZZ}

\pubyear{2017}

\hypersetup{draft} 
\begin{document}
\label{firstpage}
\pagerange{\pageref{firstpage}--\pageref{lastpage}}
\maketitle

\begin{abstract}
We investigate whether the regular Galilean satellites could have formed in the dead zone of a circumplanetary disc. A dead zone is a region of weak turbulence in which the magnetorotational instability (MRI) is suppressed, potentially an ideal environment for satellite formation. With the grid-based hydrodynamic code, FARGO3D, we examine the evolution of a circumplanetary disc model with a dead zone. Material accumulates in the dead zone of the disc leading to  a higher  total mass and but a similar temperature profile compared to a fully turbulent disc model. The tidal torque increases the rate of mass transport through the dead zone leading to a steady state disc with a dead zone that does not undergo accretion outbursts. We explore  a range of disc, dead zone and mass inflow parameters and  find that the maximum mass of the disc is around $0.001\,M_{\rm J}$. Since the total solid mass of such a disc is much lower, we find that there is not sufficient material in the disc for in situ formation of the Galilean satellites and that external supplement is required.
\end{abstract}

\begin{keywords}
methods: numerical -- hydrodynamics -- accretion, accretion discs -- planets and satellites: formation -- planets and satellites: composition 

\end{keywords}

\section{Introduction}
\label{intr}
In our solar system, regular satellites have prograde, nearly circular and nearly coplanar orbits. These regular satellite systems exist around all giant planets, including Uranus and Neptune. These are thought to have formed in a circumplanetary disc during the later stages of the formation of the central giant planet \citep{Canup2002, Canup2006, Mosqueira2003}. The formation mechanism for satellites in a circumplanetary disc may be similar to that of planets forming in a protoplanetary disc but on a smaller scale \citep{Pollack1991}. For a sufficiently massive planet that forms in a protoplanetary disc, tidal torques may open a gap within the disc \citep{Lin1987}. Gas continues to flow through the gap and a circumplanetary disc forms \citep{Artymowicz1996,Lubow1999,DAngelo2002,Bateetal2003}. Understanding the physical structure of the circumplanetary disc can help us to explain the formation of the regular satellites.

There were several circumjovian disc models that attempt to explain the formation of the Galilean satellites. \citet{Canup2002} described a gas starved disc model which allows a satellitesimal to form and condense water ice at low temperatures near the end of disc life time. However, their 1D simulations showed that the disc is orders of magnitude less massive than the minimum-mass subnebula model. The minimum-mass subnebula is a circumjovian disc that contains the minimum amount of solids necessary to build the Galilean satellites, hence the results of \citet{Canup2002} imply that there may not have enough material to form the satellites at once. Furthermore, \citet{Estrada2009} used the minimum-mass subnebula and  2D simulations to show that the circumplanetary disc may be too hot for icy satellites to form and survive in the disc. 

The tidal truncation from the Sun constrains the size of a circumplanetary disc, which is limited to be about 0.4 times the Hill radius \citep{MartinandLubow2011}. At this radius, the viscous torque balances the tidal torque from the central star that removes angular momentum at the outer edge of a circumplanetary disc. However, the Galilean satellites around Jupiter lie within a small radius and the outermost satellite, Callisto, is at 0.0345 $R_{\rm H}$. This compact configuration must be explained either by the structure of the disc or the later satellite evolution. 

To solve both problems of mass deficiency and  disc temperature, \citet{Lubow2013} suggested that a dead zone in a circumplanetary disc provides a quiescent and cool environment suitable for icy satellite formation. Material in the disc interacts through viscosity that is thought to be driven by the MRI \citep{BH1991}. This mechanism requires a sufficiently high ionisation level in the disc to operate. Protoplanetary discs are too cold and dense for the MRI to operate throughout \citep{Gammie1996} and the situation is the same in circumplanetary discs because of their relatively similar temperature but even higher densities than the local protoplanetary disc \citep{Lubow2013,Fuji2017}. The inner regions of a circumplanetary disc are hot enough to be thermally ionised. However, farther from the planet, external sources such as cosmic rays and X-rays from the central star only ionise the surface layers, leaving a dead zone at the disc mid-plane where the MRI can not operate \citep[e.g.,][]{Gammie1996, Gammie1998}. The dead zone is a quiescent region where solids can settle to the disc mid-plane \citep[e.g.,][]{Youdin2002, Youdin2007, Zsom2011}. The surface density may also increase in the dead zone due to the low viscosity. In the case of a protoplanetary disc, planetesimals may form in such a quiescent region \citep[e.g.,][]{Bai2010, Yang2018}. If the formation mechanism of regular satellites is analogous to planet formation, then satellitesimals could form in the same way in a circumplanetary disc. We note that if the dead zone viscosity is sufficiently small, however, a steady state disc may not exist \citep{MartinandLubow2013dza}.

Circumplanetary discs with dead zones may be unstable to accretion outburst cycles \citep{Lubow2013} that are similar in nature to the gravo--magneto disc instability that is thought to explain FU Orionis outbursts in young stars \citep{Armitage2001,MartinandLubow2011,Zhu2009}. As the dead zone grows in mass, the temperature increases due to increased viscous heating or gravitational instability, and the MRI may be triggered and cause a sudden increase in the disc turbulence, resulting in an accretion outburst.  Some or all of the dead zone region becomes fully turbulent for a short period of time. After the outburst, the dead zone forms again and the cycle continues. The outbursts may cease once the mass infall rate drops off. Since the outburst may lead to the accretion of the embryos of satellites on to the planet \citep{Lubow2013}, the satellites likely form after the outbursts have finished, or farther out in the disc.

Also noticeable is that there exists large difference in bulk density between inner and outer Galilean satellites, indicating that the water snowline plays an important role in their compositions. Ganymede and Callisto, the two outer satellites, contain about 50 percent ice while the two inner satellites contain much less \citep{Kuskov2005}. The temperature of the snowline is around $170\,\rm K$ if we ignore the effects of dust size and  gas density \citep{Hayashi1981,Lecar2006}. The two outer Galilean satellites were likely formed outside the snowline since they accreted more mass from water ice condensation. The partially differentiated structure of Callisto suggests that its ice never fully melted and that the snowline radius was always inside of its orbit \citep{Lunine1982,Schubert2004}. The disc temperature depends on the mass accretion rate, and for a fully MRI active disc, the accretion rate must be substantially lower than the accretion rate during the T Tauri period for the snowline radius to be in the middle of the Galilean satellites \citep{Canup2002, Estrada2009}. On the other hand, the disc model with a dead zone can have larger accretion rates for a cooler disc \citep{Lubow2013}. Thus, a comprehensive parameter study is necessary to explain the formation of Galilean satellites by considering the snowline radius. 

In this article, we extend the work of \citet{Lubow2013} who modelled a circumplanetary disc with a dead zone in 1D. We use two-dimensional simulations that allow us to properly take into account the tidal torque from the Sun. In Section 2, we describe our layered disc model for a circumjovian disc. We show simulation results from fully turbulent disc models in Section 3 and from those models with a dead zone in Section 4. In Section 5, we discuss the implications of our results and potential solutions to the formation of the Galilean satellites in the circumjovian disc. Finally, we draw our conclusions in Section 6.

\section{Circumplanetary disc model}


We model a circumplanetary disc in 2D with FARGO3D that solves the hydrodynamical equations in the cylindrical coordinate system ($R$,$\phi$) that is centred on the planet \citep{Benitez2016}. The inner boundary is at  $R_{\rm in} = 0.003\, R_{\rm H}$ and the outer boundary is at $ R_{\rm out} = 1.0\, R_{\rm H}$. $R_{\rm H}$ is the Hill radius of Jupiter

In the radial direction we take 256 grid points distributed in logarithmic scale, and in the azimuthal direction we take 128 grid points at regular intervals. We set the mass of the planet $M_{\rm p}$ to be one Jupiter mass, $\rm M_{J}$, and the coordinate system corotates with the planet. 
Hence, we put a star with mass $M =1\,\rm M_{\odot}$ at an orbital radius of $R = 14.3 R_{\rm H}=5.2\,\rm au$, which revolves the centre of the grid at an angular frequency of Jupiter's orbit. The Hill radius of Jupiter is 
\begin{equation}
R_{\rm H} = 5.2\left(\frac{M_{J}}{3 M_{\odot}}\right)^{1/3}{\,\rm au} \approx 0.36 \ \rm{au}.    
\end{equation}

The initial surface density is set to be uniform and small, $\Sigma$ = $0.001\, \rm g \, cm^{-2}$. The initial disc aspect ratio is $H/R=0.05$ everywhere. Therefore, the initial sound speed is 
\begin{equation}
c_{\rm s}=0.05\sqrt{\frac{GM_{\rm J}}{R}} \ \rm {cm\ s^{-1}},
\end{equation}
where $G$ is the gravitational constant.

The inner and outer radial boundaries have  free flow boundary conditions. The density, energy and radial velocity are copied from the last active zones to the ghost zones. Gas can only flow out through the boundary and no gas can flow into the mesh from beyond the inner and outer boundaries. Hence, the radial velocity in the ghost zones are set to be zero when it is toward the active zone. The azimuthal velocities in the ghost zones are set to their local Keplerian velocities.

The energy equation in our model reads
\begin{equation}
 \frac{\partial e}{\partial t} + {\nabla}\cdot(e v) = -p{\nabla}\cdot v + Q_{+} - Q_{-},
\end{equation} 
where $e$ is  thermal energy density (thermal energy per unit area), v is flow velocity, $p$ is vertically integrated pressure, $Q_{+}$ is a heating source term and $Q_{-}$ is a cooling source term. 
From this equation, each cell in the mesh grid gains or loses thermal energy because of flow advection, compressional work, viscous heating, and radiative cooling, respectively. The heating term can be written as \citep{collins1998,DAngelo2003}
\begin{equation}
Q_{+} = \frac{1}{2\nu\Sigma}\left( \tau^2_{\rm RR} + \tau^2_{\phi\phi} +  \tau^2_{\rm R\phi} \right) + \frac{2\nu\Sigma}{9}(\nabla\cdot v)^2,
\label{equation:heat}
\end{equation}
where $\nu$ is turbulent viscosity and $\tau_{\rm RR} , \tau_{\phi\phi}$ and $\tau_{\rm R\phi}$  are the components of the viscous stress tensor in radial-radial, azimuthal-azimuthal and radial-azimuthal directions. The cooling is determined by black body radiation near the surface of the disc
\begin{equation}
Q_-=\sigma {T_{\rm e}}^4,
\end{equation}
where $\sigma$ is the Stefan-Boltzmann constant and $T_{\rm e}$ is the effective temperature. The effective temperature is related to the mid-plane temperature, $T_{\rm c}$, by
\begin{equation}
T_{\rm c}^4=\tau T_{\rm e}^4,
\end{equation}
where the optical depth is
\begin{equation}
\tau=\frac{3}{8}\kappa \frac{\Sigma}{2},
\end{equation}
and the opacity is $\kappa= aT_{\rm c}^{b}$, where $T_{\rm c}$ is in Kelvin. In our model, we take $a = 0.02$cm$^{2}$ g$^{-1}$  and $b = 0.8$ assuming that dust dominates the absorption properties of matter in the disc \citep{Bell1994,Zhu2009}.
The mid-plane disc temperature is derived from the internal energy via
\begin{equation}
e = \frac{\Sigma {\cal R} T_{\rm c}}{(\gamma - 1)\mu},
\end{equation}
where $\gamma=5/3$ is the adiabatic index and  $\cal R$ is the gas constant. Most of the mass of the disc is in molecular hydrogen so the mean molecular weight is $\mu=2.4\,\rm g\, \rm{mol}^{-1}$ \citep{Dutrey2014}.

We adopt the layered disc model of \citet{Armitage2001} and assume a spatially varying turbulent viscosity $\nu$ that depends on the local condition of the disc \citep[see also][]{Zhu2009, MartinandLubow2011, MartinandLivio2012}. The critical surface density, $\Sigma_{\rm crit}$, and the critical disc temperature, $T_{\rm crit}$, are the two constants we use to determine if a location is fully MRI active or contains a dead zone, as discussed below.  

The disc is ionised by external sources, such as cosmic rays or X-rays. For a lower disc surface density, $\Sigma<\Sigma_{\rm crit}$, the gas is fully MRI active, where the ionising sources can penetrate deep into the mid-plane. However, the value of the critical surface density $\Sigma_{\rm crit}$ is uncertain. For discs around T Tauri stars, at a radial distance less than 0.1 au where cosmic rays ionise effectively, the critical surface density is $\Sigma_{\rm crit} \approx\, 100\, \rm g\, cm^{-2}$ \citep{Umebayashi1981}. If chemical reactions of charged particles are taken into account, this value could be lower \citep{Sano2000}. The MRI could operate in  a disc ionised by stellar X-rays with $\Sigma_{\rm crit} \approx 1 - 30\, \rm g\, cm^{-2}$ \citep{Turner2008,Bai2009}. In the circumjovian disc, \citet{fujii2014} found the MRI can only be sustained if the surface density is below $ 10\, \rm g\, cm^{-2}$ for the region around the Galilean satellites. Due to the uncertainty, we let $\Sigma_{\rm crit}$ be a free parameter that is in the range of $ 1 - 10\, \rm g\, cm^{-2}$. 

For disc temperature $\gtrsim$ 800 K, collisional ionisation of potassium makes the ionisation fraction increases exponentially with temperature \citep{Umebayashi1983} and thus we take this temperature to be the critical above which MRI can operate. In other words, when $ T > T_{\rm crit}$, the gas is fully MRI active no mater how large $\Sigma$ is.  Therefore, if a region in a disc has $\Sigma > \Sigma_{\rm crit}$ and $ T < T_{\rm crit}$, a dead zone around the mid-plane forms and the viscosity in the dead zone is much lower than that in a fully MRI turbulent region because of the inefficient transport of angular momentum.

The viscosity in the fully MRI active parts of the disc is parametrized by the \cite{SS1973}  $\alpha$ parameter
\begin{equation}
\nu=\alpha c_{\rm s}H,
\end{equation}
where the sound speed $c_{\rm s}=\sqrt{{\cal R}T_{\rm c }/\mu}$ and $H$ is the vertical scale height of the gas. Observations of FU Orionis suggested that $\alpha \approx 0.01$ \citep{Zhu2007}. Observations of X-ray binaries and dwarf novae that have a fully turbulent disc  gave an estimate of $\alpha \approx 0.1 - 0.4$ \citep{King2007,Martin2019}. MHD simulations found $\alpha \sim 0.01$ although those models depends on numerous parameters such as the net magnetic flux \citep[e.g.,][]{Hawley1995,Johansen2006,yang2009,Bai2013}, stratification 
\citep{Daviesetal2010, Bodo2014, Ryan2017} and treatments of small-scale dissipation \citep{Fromang2007, Oishi2011, Meheut2015,Walker2016}. In this work, we assume the $\alpha$ parameter in a fully MRI turbulent region to be 0.01.
In regions with a dead zone, the viscosity is approximated with
\begin{equation}
\nu=\alpha_{\rm dz} c_{\rm s}H,
\end{equation}
where the parameter $\alpha_{\rm dz}$ is much smaller than $\alpha$ but may not be zero since there are several possible mechanisms to drive turbulence in the dead zone. First, hydrodynamic instabilities such as the baroclinic instability may operate in the dead zone \citep[e.g.,][]{Klahr2003,Peretson2007,Lesur2010}. A pressure gradient over surfaces of constant density can generate vorticity leading to $\alpha \approx 5\times 10^{-3}$ \citep{Lyra2011}.  Second, shearing box simulations showed that MHD turbulence generated in the disc surface layers can penetrate into the mid-plane and exert a non-zero Reynolds stress there \citep{Fleming2003, Oishi2007,Turner2007, Oishi2009, Okuzumi2011}.  Finally, self-gravity can also produce turbulence in the mid-plane of the disc (e.g. Lodato \& Rice 2004). The turbulence driven by self-gravity can be up to $\alpha$ $\sim$ 0.1 \citep{shi2014}. However, self-gravity is unlikely to play a role in the circumplanetary disc \citep{MartinandLubow2013dza} and we do not consider this effect in our work. In any case, there are still some uncertainties for the turbulent transport in the dead zone and we set the viscosity parameter in the dead zone to be a constant in the range of $\alpha_{\rm dz} = 10^{-5}$ to $10^{-3}$.  

We note that the region of the disc with a  dead zone should ideally be modelled with a vertically varying $\alpha$ viscosity  parameter \citep[e.g.][]{Pierens2010} based on fits to MHD simulations \citep[e.g.][]{Okuzumi2011,Gressel2011,Uribeetal2011,Uribeetal2013}. However, since we use 2D simulations in $R-\phi$, we do not model the vertical structure of the disc. Instead, we consider $\alpha$ as a density-weighted, vertically integrated quantity to model the radial flow through the disc \citep[e.g.][]{Suzuki2010,Suzuki2016} and leave it as a free parameter (Section~\ref{DMDZ} ). 
This approximation is reasonable as long as the column density of the active layer is much smaller than the column density in the dead zone layer. Since the active layer is generally small \citep[e.g.][]{Martinetal2012a,Martinetal2012b,fujii2014}, this approximation does not significantly affect the disc mass in our steady-state models. In the case of a much smaller dead zone size, our models represent the upper limit to the disc mass. 

We note that density waves in the circumplantary disc can also drive disc accretion. \citet{Zhu2016} found that the tidal torque from the Sun can excite spiral density waves in the circumjovian disc and it results in shocks that transport angular momentum through the disc. Thus, the effective viscosity parameter in a dead zone may be larger than $\alpha_{\rm dz}$. Their simulations showed that the effective viscosity value is in the range $10^{-4} \leq \alpha_{\rm eff} \leq 10^{-2}$ in a circumjovian disc. In our 2D simulations, the tidal torque is self-consistently included and may produce a comparable $\alpha_{\rm eff}$ in the disc.

To model the mass accretion from the protoplanetary disc onto the circumplanetary disc, we continuously deposit gas mass at a constant rate over all values of $\phi$ at a radius of $R_{\rm add}$ = 0.33 $R_{\rm H}$ which is the radius determined by the angular momentum of the material that falls into the Hill sphere \citep{Quillen1998, Estrada2009}. We assume that the mass infall rate $\dot{M}$ is similar to that in a protoplanetary disc and hence we consider $\dot{M}$ in the range of $10^{-11}-10^{-9} \rm M_{\odot} yr^{-1}$ \citep{Bateetal2003, lubow2006, Ayliffe2009, Zhu2016} (See \citet{Lubow2012} for more discussion). 

The fully MRI turbulent disc may evolve differently compared to a disc with dead zone. Strong turbulence leads to a lower surface density and a higher temperature. Thus, it may result in satellites being hard to form in a disc.  In the next two sections we describe the results of fully MRI turbulent disc simulations and then our disc models with a dead zone. In Table 1, we summarise all of the simulation parameters.

\begin{table*}
\centering
\caption{The parameters of the simulations. Column 1 is the name of the simulation. Column 2 is the mass infall rate on to the circumplanetary disc. Column 3 is the critical surface density. Column 4 is the viscosity $\alpha$ parameter in MRI active regions. Column 5 is the viscosity parameter in the dead zone, $\alpha_{\rm dz}$. Column 6 is the simulation end time. Column 7 is the mass of the disc at the end of the simulation. Column 8 is the snow line radius at the end of the simulation.  }
\label{tab:example_table}
\begin{tabular}{ccccccccc} 
\hline
\textbf{Model} & \textbf{$\dot{M}$} & $\Sigma_{\rm crit}$ & $\alpha $ & $\alpha_{\rm dz}$ &Simulation time & Mass of disc & Snowline radius & Resolution \\
 & ($\rm M_{\odot}\, yr^{-1}$)& ($\rm g\, cm^{-2}$)& & & (yr)&  ($M_{\rm J}$) & ($R_{\rm H}$)& ($R$,$\phi$) \\
\hline
\hline
S1 &  $10^{-9}$  & - & 0.01 & - & 3000 & $2.0\times10^{-4}$ & $7.0\times10^{-2}$ & 256, 128\\
S2 &  $10^{-10}$  & - & 0.01 & - & 3000 & $4.3\times10^{-5}$ &$2.5\times10^{-2}$ & 256, 128 \\
S3 &  $10^{-11}$  & - & 0.01 & - & 3000&$5.2\times10^{-6}$ &$7.0\times10^{-3}$ & 256, 128 \\
\hline
D1 &  $10^{-10}$  & 1 & 0.01 &$10^{-4}$ & 10000&$3.9\times10^{-4}$ &$2.4\times10^{-2}$ & 256, 128\\
C1 &  $10^{-10}$  & 10 & 0.01 & $10^{-4}$ & 10000&$4.0\times10^{-4}$ &$2.5\times10^{-2}$ & 256, 128\\
A1 &  $10^{-10}$  & 1 & 0.01 &$10^{-3}$ & 10000&  $1.8\times10^{-4}$ & $2.9\times10^{-2}$ & 256, 128\\
A2 &  $10^{-10}$  & 1 & 0.01 &$10^{-5}$ & 10000&  $5.2\times10^{-4}$ & $2.7\times10^{-2}$ & 256, 128\\
M1 &  $10^{-9}$  & 1 & 0.01 &$10^{-4}$ & 4000& $1.0\times10^{-3}$&$6.8\times10^{-2}$ & 256, 128\\
M2 &  $10^{-11}$  & 1 & 0.01 &$10^{-4}$ & 17000& $1.4\times10^{-4}$&$5.7\times10^{-3}$ & 256, 128\\
H1 &  $10^{-10}$  & 1 & 0.01 &$10^{-4}$ & 6300&$3.9\times10^{-4}$ &$2.3\times10^{-2}$ & 512, 256\\

\hline
\label{table1}
\end{tabular}
\end{table*}

\section{Fully MRI turbulent disc models}

\begin{figure}
\centering
\includegraphics[width=8.3cm]{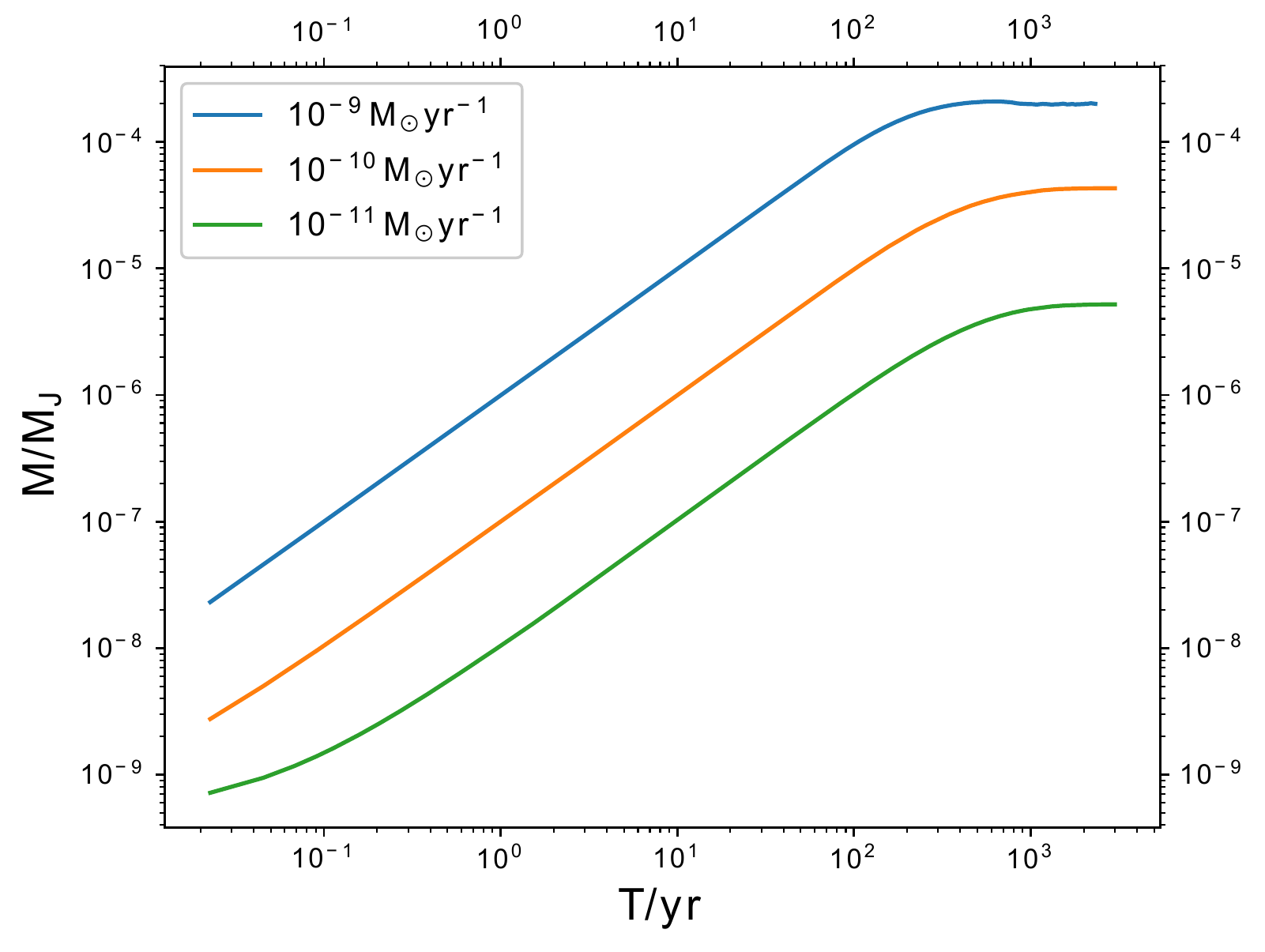}
\caption{The total mass of the fully MRI turbulent disc models S1($\dot M = 10^{-9}\,\rm M_{\odot}\, yr^{-1}$), S2($\dot M = 10^{-10}\,\rm M_{\odot}\, yr^{-1}$) and S3($\dot M = 10^{-11}\,\rm M_{\odot}\, yr^{-1}$) as a function of time.}
\label{ss2}
\end{figure}

\begin{figure*}
\centering
\includegraphics[width=8.3cm]{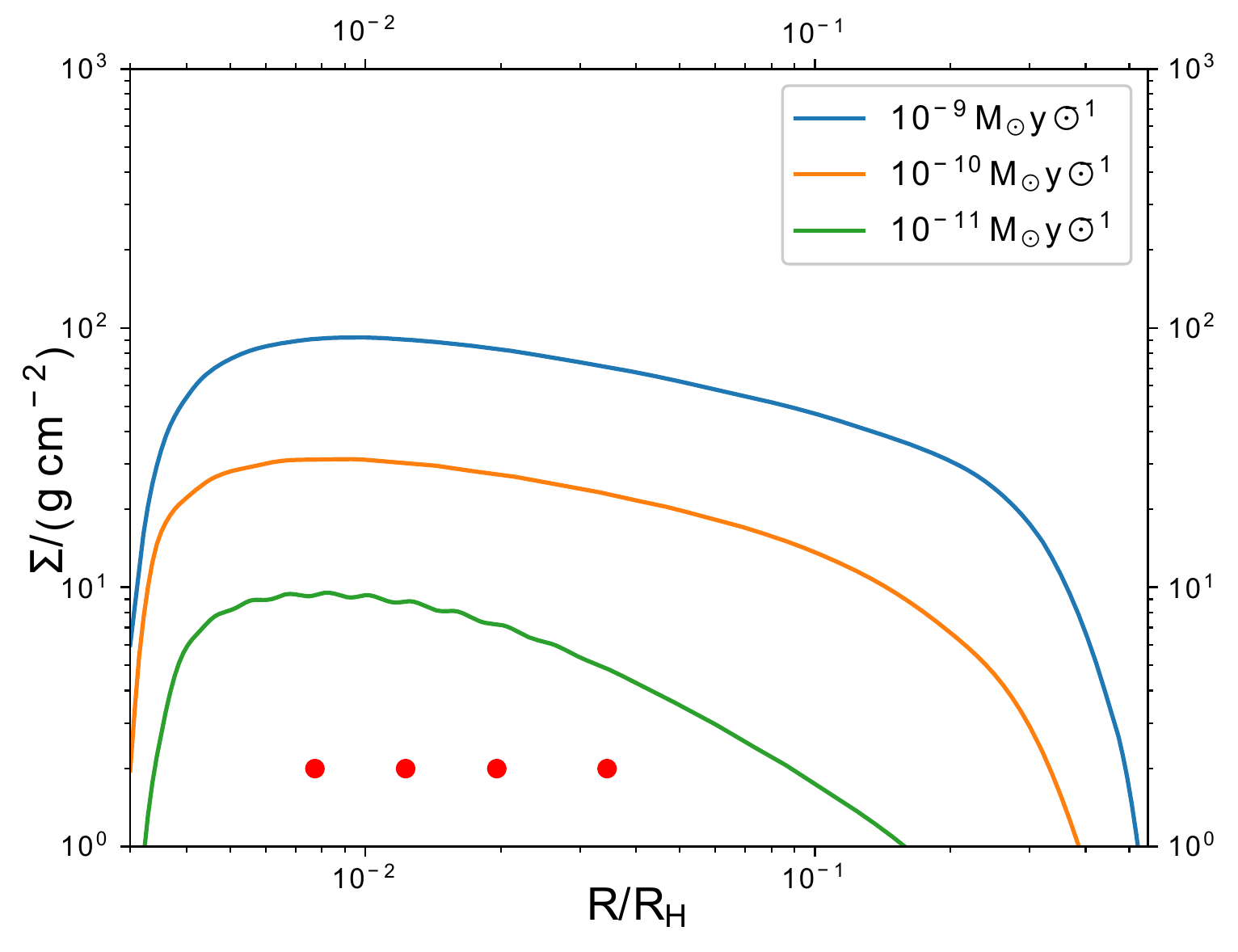}
\includegraphics[width=8.3cm]{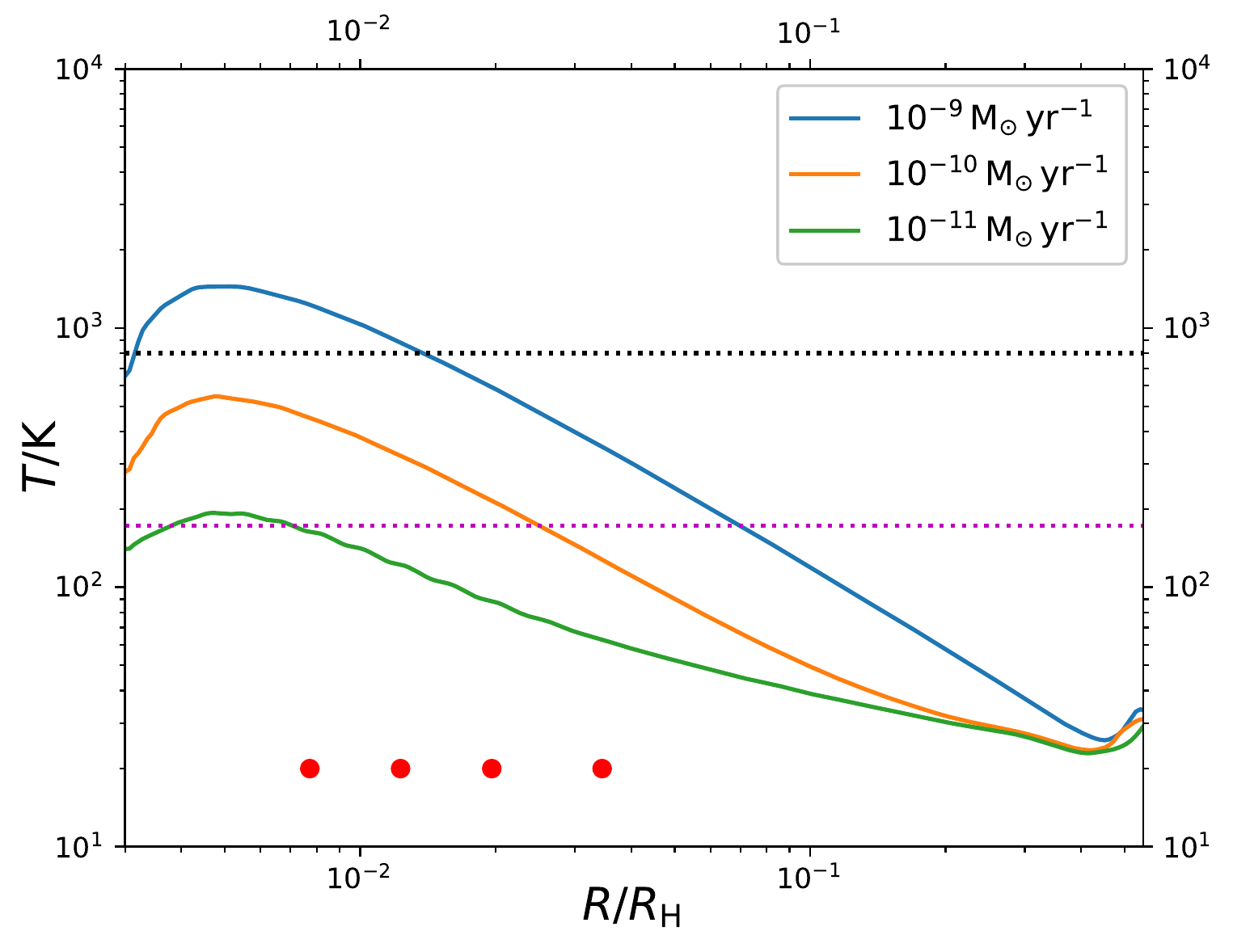}
\caption{Steady-state, fully MRI turbulent circumplanetary disc models with mass infall rates $\dot M=\,\rm 10^{-9}\,$ (blue lines),   $ 10^{-10}$ (orange lines) and $\rm 10^{-11}\, M_{\odot}\, yr^{-1}$ (green lines) at time $t$ = $3000 \,\rm yr$.
Left panel: Surface density. Right panel: Temperature. The black dotted line is the critical temperature $T_{\rm crit}= 800\,\rm K$ and the purple dotted line is the snowline temperature, $T_{\rm snow}=170\,\rm K$. The four red dots represent the orbital locations of Io ($R$ = 0.0077 $R_{\rm H}$), Europa ($R$ = 0.0123 $R_{\rm H}$), Ganymede ($R$ = 0.0196 $R_{\rm H}$) and Callisto ($R$ = 0.0345 $R_{\rm H}$). }
\label{ss}
\end{figure*}

\begin{figure}
\centering
\includegraphics[width=8.3cm]{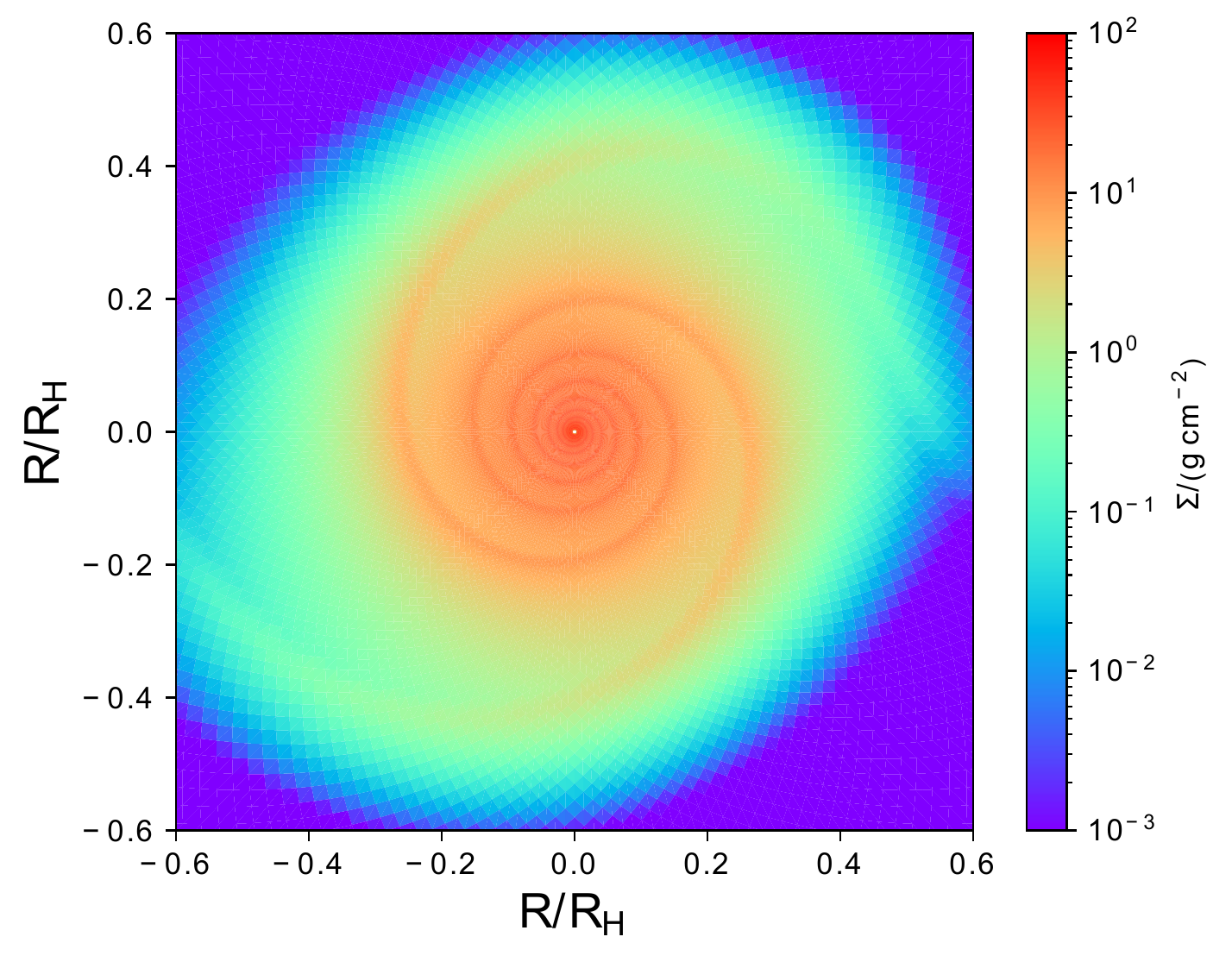}
\caption{Surface density of steady-state, fully MRI turbulent disc at time  $t = 3000\, \rm{yr}$ (S2 model with infall accretion rate $\dot M=10^{-10}\,\rm M_\odot \, yr^{-1}$). The sun is 5.2 au away (14.3 $R_{\rm H}$) on the right hand side of the figure.} 
\label{ss2D}
\end{figure}

First we consider fully MRI turbulent disc models with $\alpha=0.01$ everywhere. Fig.~\ref{ss2} shows the total mass of each disc as a function of time for three different infall accretion rates, $\dot M = 10^{-9}\,\rm M_{\odot}\, yr^{-1}$ (model S1), $\dot M=10^{-10}\,\rm M_{\odot}\, yr^{-1}$ (model S2) and $\dot M= 10^{-11}\,\rm M_{\odot}\, yr^{-1}$ (model S3). Each model is run until it reaches a steady state. The disc initially builds in surface density due to the mass accretion from the outer disc. The disc temperature also increases due to viscous heating (Eq.~\ref{equation:heat}). The viscosity increases with increasing disc temperature and the disc spreads outwards until the disc reaches a quasi-steady state.

Fig.~\ref{ss} shows the surface density and temperature profiles at time $t=3000\,\rm  yr$. The disc spreads outwards but around a radius of about $0.4\,R_{\rm H}$, where the tidal torque from the Sun exceeds the viscous torque. Consequently, the tidal torque truncates the disc there \citep[see also][]{MartinandLubow2011}.

These simulations show that a fully MRI turbulent disc model is inadequate to explain the formation of the Galilean satellites. The higher the infall accretion rate, the larger the mass of the steady disc. For the highest mass infall rate $\dot M = 10^{-9}\,\rm M_{\odot}\, yr^{-1}$ (model S1), the total gas mass of the disc is comparable with the total mass of the Galilean satellites($\sim 2\times10^{-4}M_{\rm J}$). The solid mass density in the disc is much smaller than the mass of the satellites since
the dust-to-gas ratio in a protoplanetary disc is only 1 - 10\% \citep[e.g.][]{Soon2019}. Dust drifts from the outer gap region of a protoplanetary disc and the positive gas pressure gradient produced by the gap dams most of the pebble-sized dust. Because most dust has already grown to the size of pebbles when it reaches the region around gas giants, it is dammed at the outer edge of the gap \citep{Lambrechts2012,Okuzumi2012,Sato2016}. Thus, the dust-to-gas mass ratio is thought to be < 1 \% in a circumplanetary disc \citep{Adachi1976,Zhu2012}. Furthermore, the disc temperature is above the snowline temperature within the region of the Galilean satellites and therefore the temperature is too hot to explain the formation of icy satellites.

On the other hand, for the lower mass infall rates $\dot M \lesssim 10^{-10}\,\rm M_\odot \, yr^{-1}$  (models S2 and S3), the temperature around the orbital location of Callisto is below the snowline temperature, and water ice may condense there. However, the total masses in these two models ($4.3 \times10^{-5}M_{\rm J}$ and $5.2\times10^{-6}M_{\rm J}$)  are not high enough for the Galilean satellites to form and grow to their current masses.

Fig.~\ref{ss2D} shows the surface density of model S2 in 2D at time $t=3000\,\rm yr$. Our models show prominent density wave structures in the circumplanetary disc that are not present in the 1D models of \citet{Canup2002} and \citet{Lubow2013}. The density waves are excited in the vicinity of the Lindblad resonance due to the tidal torques from the Sun, and torques are exerted on the disc at radii where the waves damp \citep{Goldreich1979}.

Because the fully MRI turbulent disc is either not sufficiently massive or too hot to explain the formation of the Galilean satellites in situ, we now consider disc models with dead zones that allow a high surface density to accumulate while remaining cool enough for icy satellite formation. 

\section{Disc models with a dead zone}

In this Section we first consider the evolution of a fiducial disc model with a dead zone and then we vary different disc parameters to understand their effects on the disc evolution and satellite formation. 

\subsection{Fiducial disc model with a dead zone}

\begin{figure*}
\centering
\includegraphics[width=8.3cm]{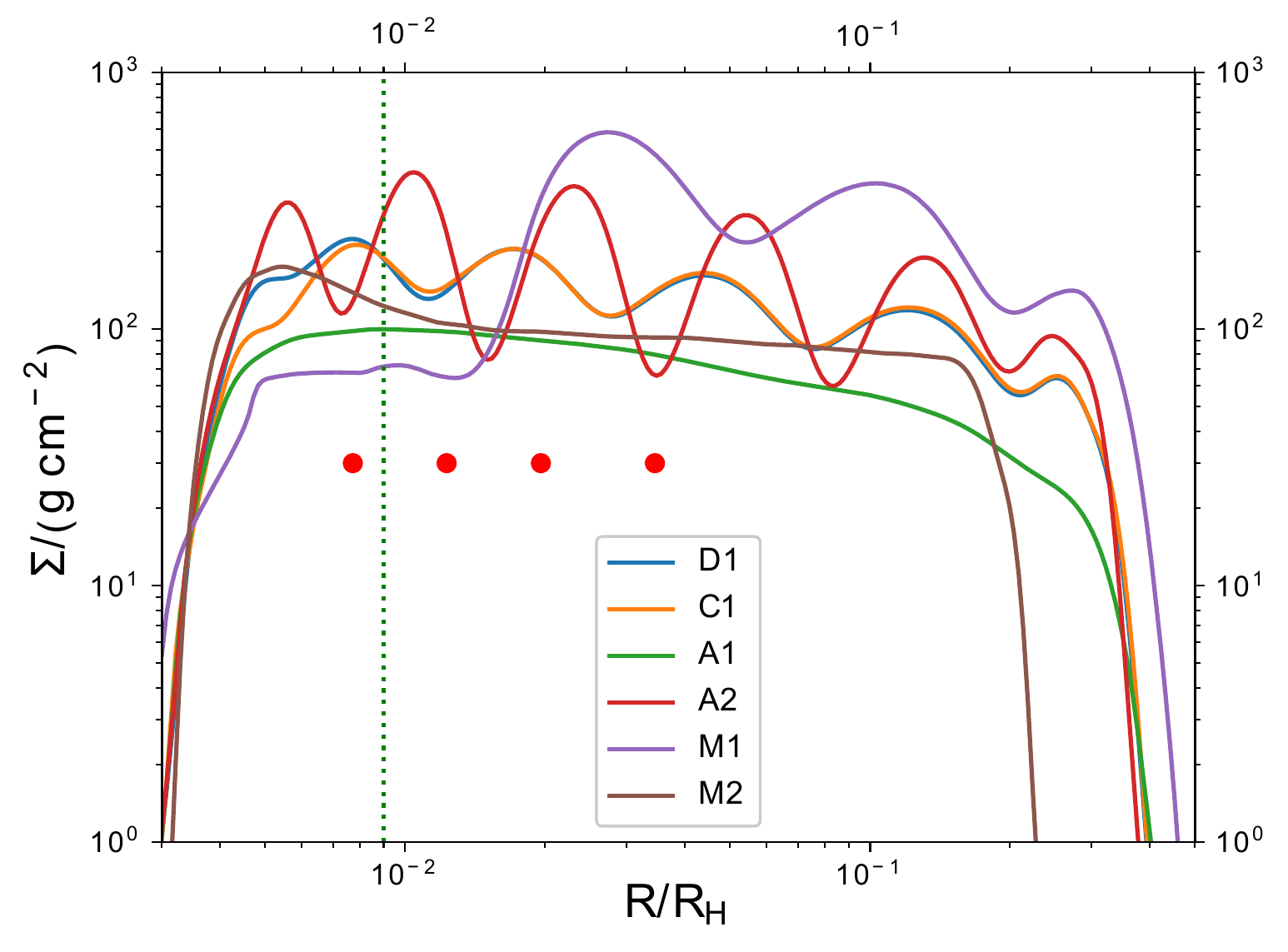}
\includegraphics[width=8.3cm]{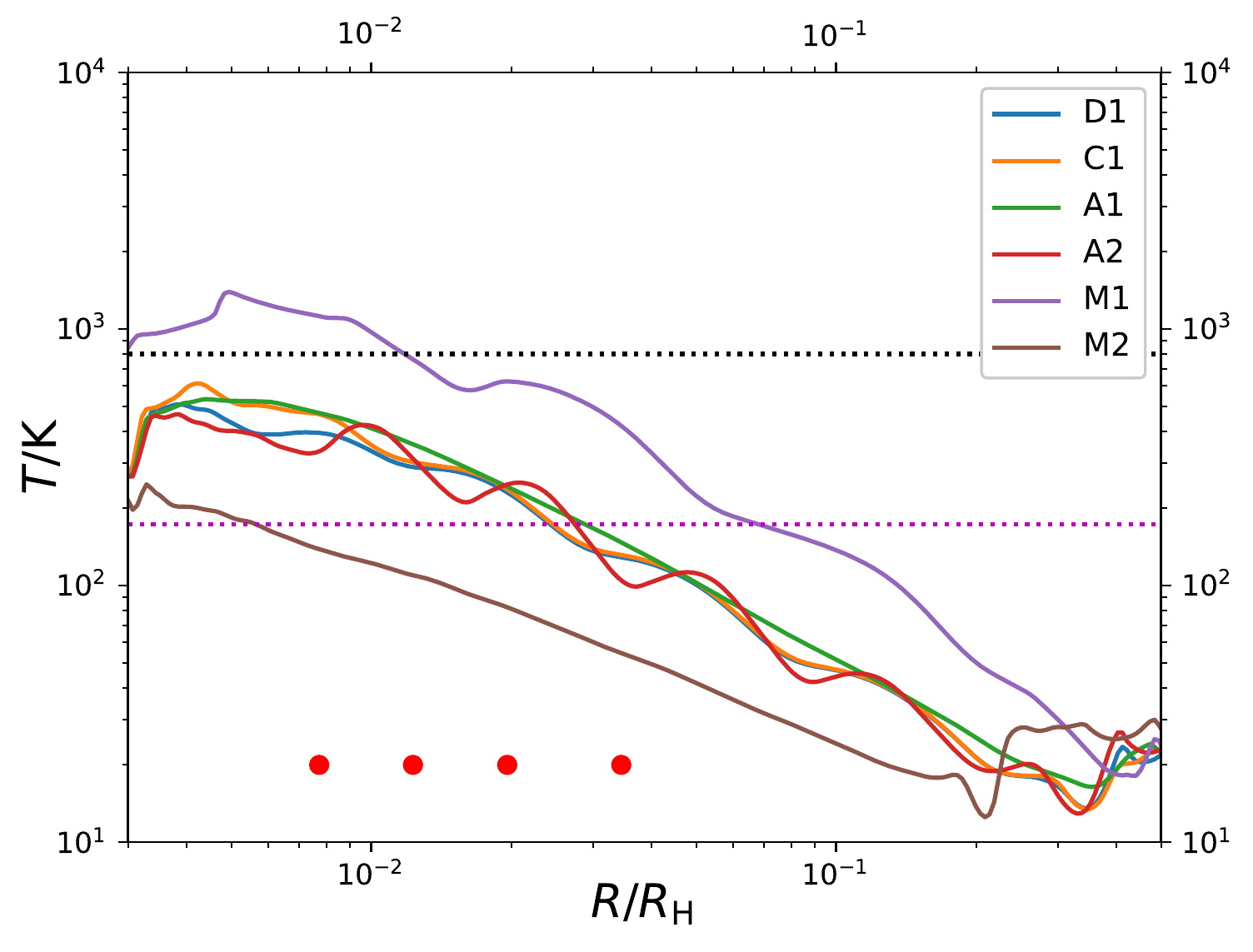}
\caption{Left: Surface density of the models with a dead zone. Right: Disc temperature of the models with a dead zone. The horizontal black dotted line is the critical temperature $T_{\rm crit}= 800\,$K and the horizontal purple dotted line is the snow line temperature, $T_{\rm snow}= 170\,\rm K$. The red dots show the location of the four Galilean satellites.
The snapshots are shown at 10000 yr for D1, C1, A1 and A2, 3800 yr for M1 and 170000 yr.}
\label{DDTT}
\end{figure*}

\begin{figure}
\centering
\includegraphics[width=8.3cm]{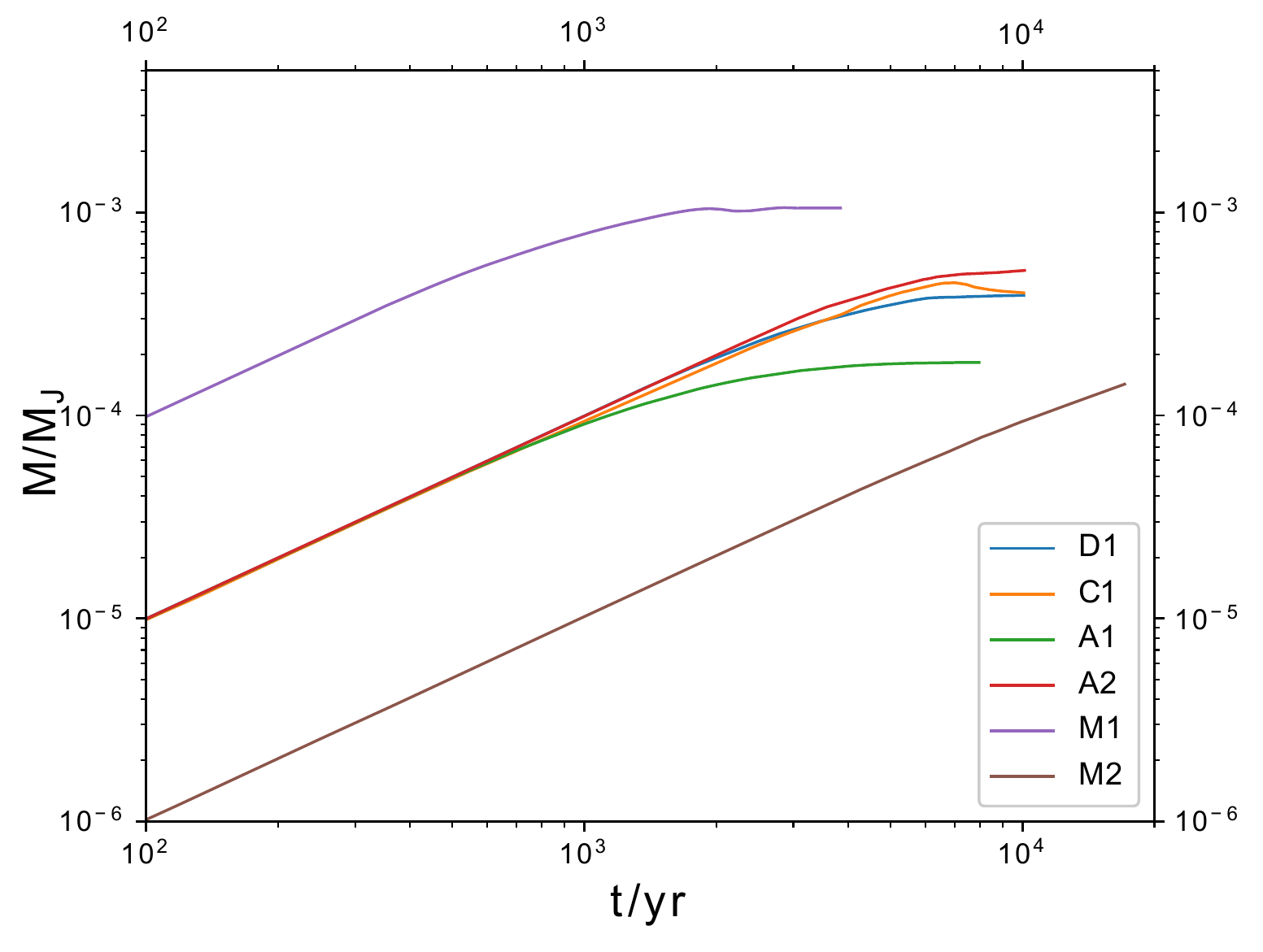}
\caption{The total mass of the disc for dead-zone models D1, C1, A1, A2, M1 and M2 as a function of time.}
\label{MM}
\end{figure}

We use a fiducial model (model D1) with the same parameters as model R9 in \citet{Lubow2013} except $\alpha_{\rm dz} = 10^{-4}$ and $\Sigma_{\rm crit} = 1\,\rm g\,{cm}^{-2}$. The mass infall rate is $\dot{M} = 10^{-10}\, \rm M_{\odot}\, yr^{-1}$ and $\alpha = 0.01$. The blue lines in  Fig.~\ref{DDTT} show the surface density and the temperature profiles of the disc at time $t = 10000\,\rm  yr$. The blue line in Fig.~\ref{MM} shows the evolution of the total disc mass in time. The dead zone builds up in mass and expands outwards. Due to the tidal torque, the disc is truncated around $0.4\, R_{\rm H}$. The steady state mass of the disc is almost an order of magnitude larger than that of the fully MRI active disc with the same infall accretion rate (model S2), but the temperature of the disc remains similar to that in model S2. 

The tidal torque increases the effective viscosity in the dead zone. Due to the density waves, the disc forms five local peaks in surface density. More specifically, those peaks are formed due to some nonlinear properties of the spirals which are related to the m=2 lindblad resonances. They roughly correspond to where the torque density is highest, but themselves are not spirals. Therefore, the depth of the density peak has no direct association with the intensity of the spiral.

Satellitesimals may be trapped and grow in the peaks. The temperatures of two outer peaks are below the snowline temperature and provide an ideal environment to form Ganymede and Callisto with abundant water ice. Eventually, satellitesimals at the outer peaks migrate inward to the current locations of Ganymede and Callisto. On the other hand, the two inner peaks have temperatures above the snowline temperature.  This may explain why Io and Europa contain no water ice or only a little water ice. Thus, the temperature difference between inner peaks and outer peaks can result in the different compositions for the inner and outer Galilean satellites. 

The total mass of the disc approaches a steady state value. Thus, even though there is a dead zone within the disc, material is able to flow through it because of the effective viscosity driven by the tidal torques. Material does not continue accumulating in the dead zone as it did in the 1D models of \citet{Lubow2013}. There is a limit to the amount of mass that the circumplanetary disc may contain even with a dead zone. In this model we do not expect outbursts to occur where the MRI is triggered in the dead zone since there is no local peak in the temperature profile that can reach the critical temperature.  The total mass of the steady state gas disc is comparable to the total  mass of the Galilean satellites (about 4 $\times$ $10^{-4} \rm M_{J}$). 

\subsection{Effect of the critical surface density}

In this section, we consider how the critical surface density below which the gas is sufficiently ionised by external sources for the MRI to operate affects the circumplanetary disc evolution. We run a simulation with the same parameters as our fiducial disc model but with a higher critical surface density $\Sigma_{\rm crit} = 10\rm \,g\,{cm}^{-2}$. The orange lines in Fig.~\ref{DDTT} and in Fig.~\ref{MM} show the surface density and the disc temperature profiles in steady state and the evolution of total disc mass of model C1. The surface density is similar to that of the fiducial model D1 because the surface density in the steady state is much higher than the critical surface density and the dead zone is largely unaffected. Therefore, the total mass of the disc is similar to model D1. The innermost region of the disc ($r \leq 10^{-2} R_{\rm H}$) has a higher disc temperature than in model D1 and is slightly depleted, but most of the disc is similar. Thus, the critical surface density has little influence for the circumplanetary disc evolution unless the critical surface density is too high for a dead zone to form in a circumplanetary disc.

\subsection{Effect of the viscosity in the dead zone $\alpha_{\rm dz}$}
\label{DMDZ}
We now explore the effect of changing the viscosity in the dead zone. This viscosity may be induced by hydrodynamic instabilities or perturbations driven by the surface turbulent layers, as discussed in Sec~\ref{intr}.

\subsubsection{Higher $\alpha_{\rm dz}$}

Model A1 has the same parameters as the fiducial model except for a larger $\alpha_{\rm dz}$ = $10^{-3}$. The green lines in Fig.~\ref{DDTT} and in Fig.~\ref{MM} show the surface density and the disc temperature profiles in steady state and the evolution of the total disc mass of model A1. The surface density and the disc mass are lower than the fiducial model by a factor of about 2.5 due to the larger $\alpha_{\rm dz}$ that allows material to be transported efficiently and prevents the disc surface density from building up. In addition, there is no density bump in this model as a result. The temperature profile is similar to that in model D1 but is smoother.
The temperature around Callisto is below the snowline temperature but the disc around Ganymede has higher temperature than the snowline temperature.  For Ganymede and Callisto, they could form in the outer dead zone where abundant water ice exists and then migrate inward to their current locations. However, the disc mass in this model is much too low for the satellites to form.

\subsubsection{Lower $\alpha_{\rm dz}$}
Model A2 has the same parameters as the fiducial model D1 except a smaller $\alpha_{\rm dz}$ = $10^{-5}$. The red lines in Fig.~\ref{DDTT} and in Fig.~\ref{MM} show  the surface density and the disc temperature profiles in steady state and the evolution of the total disc mass of model A2. There are six density bumps in the surface density profile and the amplitudes of them are more prominent than those in the fiducial model. The structure is similar to low disc accretion rate models in \citet{Zhu2016} which are inviscid (see thier figure 5).  Due to the smaller $\alpha_{\rm dz}$, the total mass of the disc is  $5.2\times 10^{-4}\ M_{\rm J}$  and is slightly higher than the fiducial model. The temperature profile is similar to the fiducial model even though the disc has a smaller $\alpha_{\rm dz}$ because the effective $\alpha$ viscosity driven by the density waves is comparable to $10^{-4}$. The three outer peaks are below the snowline temperature. Thus, satellitesimals may be trapped and grow in those peaks  which may be an ideal environment for ice satellites. However, the total mass of the disc remains too small for the disc to have sufficient solid mass to form the satellites.

\subsection{Effect of the mass infall rate}
\subsubsection{Higher mass infall rates}

Model M1 has the same parameters as our fiducial disc model but with a higher mass infall rate of $\dot{M} = 10^{-9}\,\rm M_{\odot}\,{yr}^{-1}$. The purple lines in Fig.~\ref{DDTT} and in Fig.~\ref{MM} show the surface density and the disc temperature profiles in steady state and the evolution of the total disc mass of model M1. The disc reaches a steady state after a time of about $1800 \,$yr. The total disc mass is almost an order of magnitude higher than the fiducial model and about three that of the Galilean satellites. The dead zone forms farther out than the fiducial model, outside of the orbit of Europa. There are three density bumps in the dead zone. However, the temperature is too high for water ice to condense around Ganymede and Callisto in the innermost density bump. If the two outer Galilean satellites can form in the outer two peaks, they might be able to condense water ice there since the disc temperature there is lower than that at their current locations. This scenario is similar to model A of the Galilean satellites formed by pebbles in \citet{Shibaike2019} in which  three outer Galilean satellites formed in order at the disc radius $0.065 \,R_{\rm H}$ and then migrated inwards to their current locations. The second density bump is close to the snow line temperature. It may help to explain the melted structure of Ganymede. They would then migrate inward to their current locations.

There are no accretion outbursts even in the disc with high mass infall rate, in contrast to previous 1D simulations \citep{Lubow2012}. The outburst can only occur if sufficient material is able to build up in the dead zone to drive viscous heating and trigger the MRI. However, because the tidal torque drives an effective viscosity in the dead zone, the disc reaches a steady state rather than continues to increase in mass. 

\subsubsection{Lower mass infall rates}
Model M2 has the same parameters as the fiducial model except a lower mass infall accretion rate of $\dot{M}=10^{-11}\,\rm M_{\odot}\, yr^{-1}$. The brown lines in Fig.~\ref{DDTT} and in Fig.~\ref{MM} show the surface density and the disc temperature profiles near the end of the simulation and the evolution of the total disc mass of model M2. Because the disc is still building up at 17000 yr, the surface density profile and total disc mass are lower than fiducial model by a factor of 3. The disc temperature is much lower than the fiducial model and the whole region around Galilean satellites is below the snowline temperature. However, extrapolating the brown line in Fig.~\ref{ss2}, it takes about $10^{5}$ yr for the disc to reach the mass of the disc of model D1. Due to the longer timescale of build up of the disc, it is even more difficult to explain the formation of Galilean satellites in this model.

\subsection{Effect of the resolution}
\begin{figure}
\centering
\includegraphics[width=8.3cm]{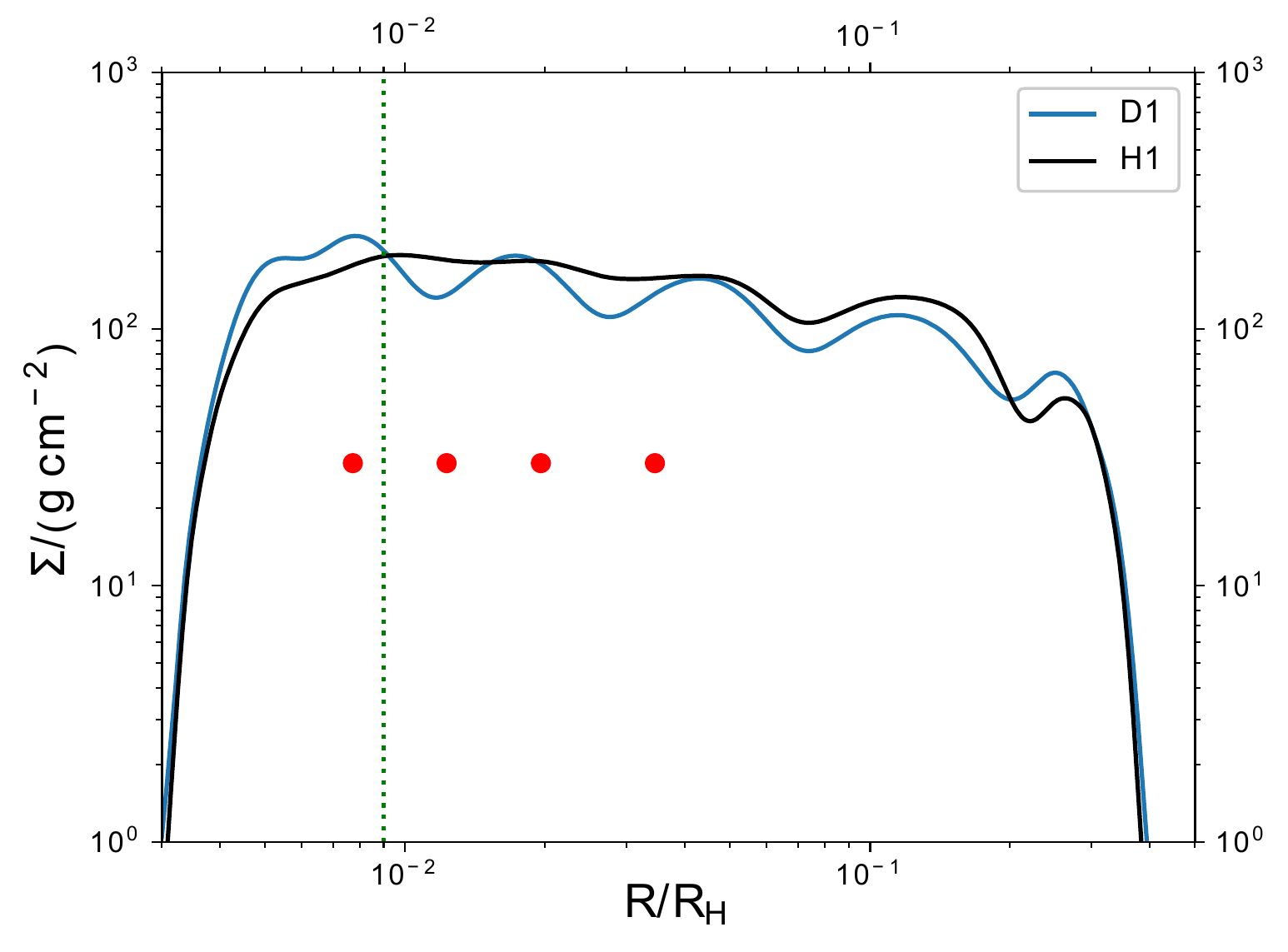}
\includegraphics[width=8.3cm]{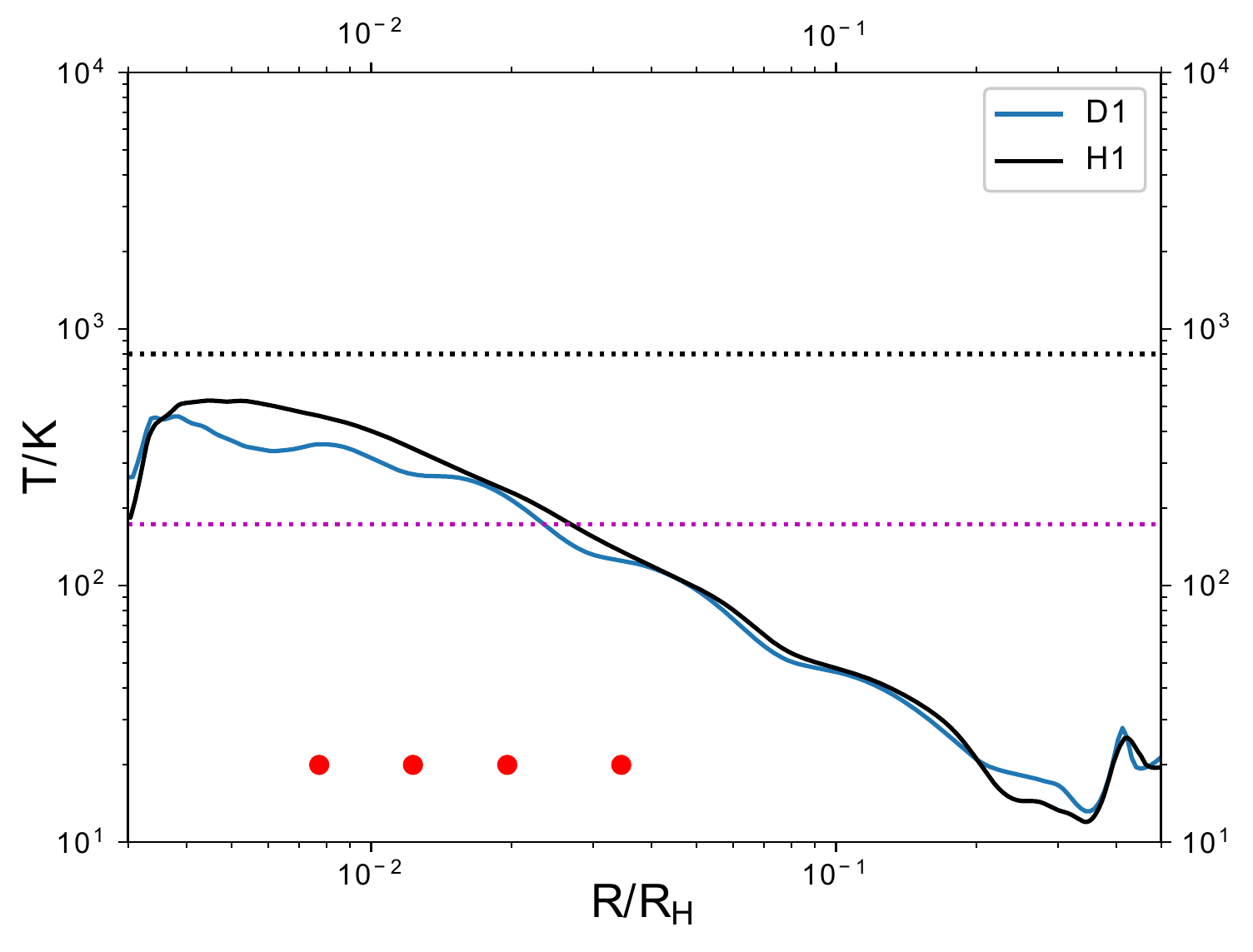}
\includegraphics[width=8.3cm]{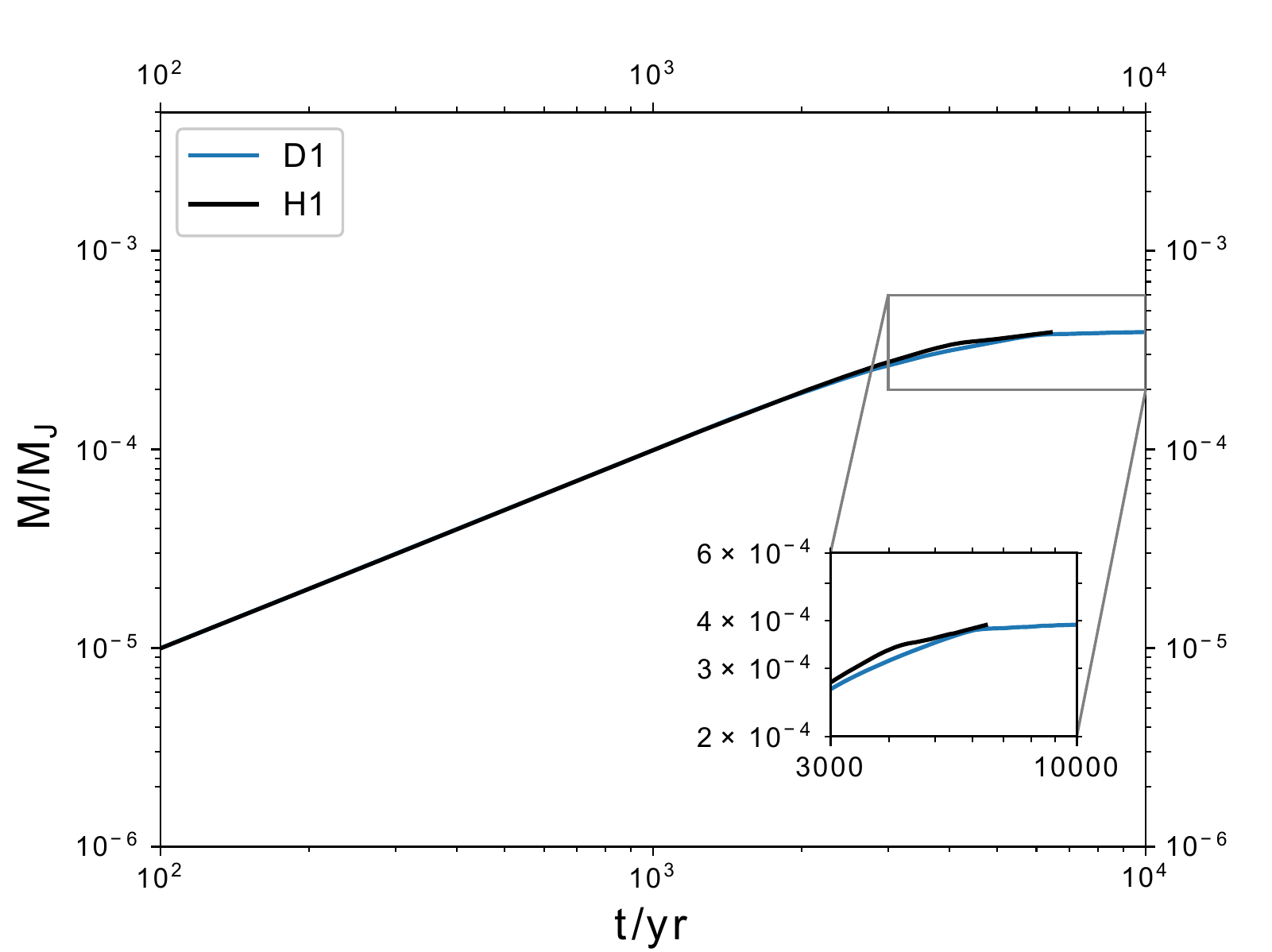}

\caption{Upper and middle panels: The surface density and disk temperature of models D1 and H1 similar to Fig.~\ref{DDTT} except that they are shown at the same time of $t$ = 6300 yr. Lower panel: The total mass of the disc for dead-zone models D1 and H1 as a function of time.}
\label{FD}
\end{figure}

Numerical viscosity could modify the accretion rate on to the disc significantly \citep[e.g.][]{Kley1999,Yasuhiro2005}. To investigate the effect of the resolution, we run the fiducial model (D1) with double the resolution in both $R$ and $\phi$ in model H1, with otherwise the same parameters. Because this simulation is more expensive than others, we just run it to $t$ = 6300 yr. To make a clear comparison, the upper two panels of Fig.~\ref{FD} show the density and temperature profiles of models D1 and H1. They are shown at the same instant of time $t$ = 6300 yr. The surface densities of the two models are similar to each other except that density bumps of model H1 are slightly shallower than model D1. The temperatures of the two models are also similar to each other but the temperature of model H1 is slightly higher than model D1 within $R$ $\simeq$ 0.02. The lower panel of Fig.~\ref{FD} shows the evolution of the total disc mass of models D1 and H1. There is a turning point at $t$ $\sim$ 4000 yr on the black line, indicating that the disc of model H1 begins to reach a quasi-steady state. It appears that the evolution of the total disc mass of model H1 converges to that of model D1 when extrapolated with time. Thus, despite using a low resolution in the outer disc in our simulations, the effect of numerical viscosity is not significant and the results from our steady-state models should be representative.

\section{Discussion}

In this work we have modelled the evolution of a circumplanetary disc and drawn some conclusions about satellite formation. We have found that a circumplanetary disc model with a dead zone does not contain sufficient material to form the Galilean satellites all at once. \cite{Canup2002} suggested that a solution to this problem is that the solid material does not need to be present all at once, but needs to be processed through the disc over time. Hydrodynamical simulations suggest that the material that flows on to the circumplanetary disc comes from the upper layers of the protoplanetary disc \citep{Machida2008, Tanigawaetal2012,Morbidelli2014, Szulagyietal2014}. These layers are depleted in large dust grains \citep{Paardekooper2006,Paardekooper2007,Birnstiel2011} and thus it is difficult to build up sufficient solids to form the Galilean satellites in this way. 

Furthermore, we have not included satellite-disc interactions that should be taken into account when satellitesimals form and grow in the circumplanetary disc \citep{Ward1997,LubowIda2010}. In the high surface density disc, the orbit of the growing satellitesimal may undergo type-I or type-II migration by satellite-disc interactions \citep{Goldreich1980}. The type-I migration time-scale can be quite short for the Galilean satellites, around $10^2$ yr \citep{Canup2002}. However, in a recent study, the Galilean satellites can lock into mean motion resonances during their migration and their orbits become stable after $10^4$ yr \citep{Moraes2018}. Besides, the dead zone allows a lower-mass object to open a gap which leads to a much slower type-II migration, because the viscous gap-opening criterion is proportional to $\alpha$ \citep{Goldreich1980}. Moreover, satellitesimals may have much longer time scales of type-I migration due to uncertain corotation torque \citep[see, e.g.,][]{Baruteau2014}. For Callisto, to release the gravitational binding energy to prevent the interior temperature from heating to the water sublimation temperature, the migration time scale should be > $10^{5}$ yr \citep{Canup2002}.

Lindblad resonances may play a role when a satellite opens a gap in the disc \citep{Canup2002, Lubow2013}. Once a planet opens a gap, the excitation of the eccentricity of the planet due to the first-order Lindblad resonance has been widely studied \citep{Goldreich2003, Ogilvie2003} and this resonance is sensitive to how clean and large the gap is. Nevertheless, at late times when the accretion rate falls off, $\dot{M}\lesssim 10^{-11}\,\rm M_{\odot}\,yr^{-1}$, the surface density around the Galilean satellites may drop to less than the critical density, thus the disc becomes fully MRI turbulent and masses of satellitesimals may not reach the gap-opening criterion. The available gas at this stage may be sufficient to damp eccentricities developed \citep{Lubow2013} and the evidence that the Galilean satellites have low eccentricities (<0.01) show those satellites have never undergone large excitations. Moreover, to prevent the high temperature around the innermost region of the dead zone, the Gallilean satellites may form farther out than their current locations and then migrate inwards later on.  In future work, we will use $N$-body simulations to test whether satellitesimals can survive in different models and migrate inward to their current locations.

\citet{Gressel2013} used 3D global hydrodynamic simulations and found that a circumplanetary disc may be tilted to the orbital plane of the planet by up to about 10$^\circ$. 
A tilted disc may affect the accretion of material on to the circumplanetary disc. However, in our simulations we have added the material at a radius corresponding to its angular momentum and thus the location of the mass deposition is unaffected by the disc tilt. Accretion of material may still be in the plane of the orbit of the planet and thus accretion causes a damping of the tilt.  Another difference with a tilted disc is that the tidal torque decreases with tilt. The torque is weakened by a factor of about two for a disc inclination of $i= 30^\circ$ and by a factor of about 20 for $i=90^\circ$ \citep{Lubow2015}. Therefore, a misaligned circumplanetary disc tends to have a larger radius than an aligned disc \citep[see also][]{Miranda2015}. This may lead to a slight increase in the mass of the circumplanetary disc. We do not expect significant warping to occur in a circumplanetary disc since the sound crossing timescale is much less than the nodal precession timescale \citep[e.g.][]{Larwoodetal1996,Larwoodetal1997,Martinetal2014b}.
The tilt of the disc may be higher for lower mass gap-opening planets \citep{Gressel2013} and their circumplanetary disc may be unstable to tilting as a result of the tidal interaction in the absence of accretion on to the disc \citep{Martin2020}.
Since the Galilean satellites are nearly coplanar to Jupiter's orbit around the Sun, it is likely that the  circumplanetary disc around Jupiter was close to coplanar.

The mass accretion rate of a circumplanetary disc can be used to predict the radiation flux of a circumplanetary disc. With mock observations, ALMA may detect circumplanetary discs of Jupiter-mass planets at high mass infall rates \citep{Szulagyietal2018,Zhu2018}. The low mass infall rate models in \citet{Lubow2012} showed that the scale of mass accretion rates are four or five orders of magnitude higher during the outburst phase. However, our models, which include a dead zone and the effect of the tidal torque that drives additional mass transport,  show that circumplanetary discs are quite stable and reach a steady state. Thus, if  accretion outbursts occur in circumplanetary discs there may be some other mechanism driving them \citep[see][]{Brittain2020}.

\section{Conclusions}

We have investigated whether a circumplanetary disc with a dead zone is able to build up sufficient mass for the Galilean satellites to form while remaining sufficiently cool for icy satellites to form. We find that a disc with a dead zone has a higher surface density and similar temperature than a fully MRI turbulent disc. We have modelled the disc evolution and varied parameters including the mass infall rate, the critical surface density required for the dead zone and the viscosity parameter. Material piles up in the dead zone but reaches a steady state that does not undergo accretion outbursts.  We find that the tidal torque from the Sun drives an effective viscosity in the dead zone that increases the mass transport rate there.  Even with a dead zone, the maximum circumplanetary disc mass reached is around $0.001\,M_{\rm J}$. 
The disc mass is  orders of magnitude smaller than the minimum-mass subnebula model in \citet{Mosqueira2003}, but consistent with the gas-starved satellite formation model \citep{Canup2002,Ward2010} and two-dimensional hydrodynamical simulations with radiative cooling in \citet{Zhu2016} that have surface densities in the range of $10 - 1000\rm g \, cm^{-2}$. Thus the circumjovian disc may not contain sufficient solid material to form the Galilean satellites at any given instant of time. We suggest that solid material must be delivered to the disc rather than in situ formation of satellitesimals and satellites \citep[e.g.][]{Ronnet2018}.

\section*{Acknowledgements}

The majority of the simulations were run on the Cherry Creek cluster at the UNLV National Supercomputing Institute. This work used the Extreme Science and Engineering Discovery Environment (XSEDE) Stampede2 at the Texas Advanced Computing Center (TACC) through allocation AST130002. CC acknowledges support from a UNLV graduate assistantship. CC thanks Ming-Tai Wu in developing the parallel function on the Cherry Creek cluster and is grateful to Pin-Gao Gu and Min-Kai Lin for helpful comments and inspiring conversations. CCY is grateful for the support from NASA via the Emerging Worlds program (Grant $\#$80NSSC20K0347).  We acknowledge support from NASA TCAN award 80NSSC19K0639. 

\section*{Data availability}
The data underlying this article will be shared on reasonable request to the corresponding author.

\bibliographystyle{mnras}
\bibliography{main}

\bsp    
\label{lastpage}
\end{document}